\documentclass[prb,twocolumn,floatfix,superscriptaddress]{revtex4}

\usepackage[dvips]{graphicx}
\usepackage{epsfig}

\usepackage{verbatim}

\usepackage{times}

\usepackage{amsmath}

\usepackage{natbib}

\newcommand{\angleb}[1]{\langle #1 \rangle}
\newcommand{\nod}{\noindent}
\newcommand{\half}{\frac{1}{2}}

\begin{document}

\title{A memory function analysis of non-exponential relaxation in
viscous liquids.}

\date{\today}


\author{Nicholas P. Bailey}
\email{nbailey@ruc.dk}
\affiliation{DNRF centre  ``Glass and Time,'' IMFUFA, Department of Sciences, 
Roskilde University, Postbox 260, DK-4000 Roskilde, Denmark}

\begin{abstract} 
We analyze data from simulations of two- and three-dimensional
(2D and 3D) glass-forming liquids using a correlation function
defined in terms of a memory function with a negative inverse power-law tail.
The self-intermediate function and the autocorrelation functions of pressure
and shear stress are analyzed; the obtained fits are very good, at least as 
good as with a stretched exponential. In contrast to the stretched exponential,
the key shape parameter---the exponent of the power-law tail---seems to be
the same for all three correlation functions. It decreases from a value 
around 2 at high temperature to a value close to 1.58 (2D), 1.50 (3D)
 at low temperatures. At 
the same time the amplitude of the tail increases towards towards a limiting 
value corresponding to a diverging relaxation time, which is related to
anomalous diffusion.
On the other hand, careful analysis of the long time behavior in the case of
the intermediate scattering function
suggests that the memory function is
cut-off exponentially, which avoids the divergence of the relaxation time.
Repeating the fits with an exponential cut-off included indicates that the
power-law exponent is in fact independent of temperature and close to 1.58/1.50
over the whole range. Instead of the divergence, a fragile-to-strong crossover 
in the dynamics, estimated to occur around $T=0.40$ for the 3D Kob-Andersen 
system. Another key parameter of the fitting procedure may be
interpreted as a short-time rate, the amount of decorrelation that occurs in
a fixed, relatively short  time interval (compared to the alpha time). This
quantity is observed to have a near-Arrhenius temperature dependence, while
its wavenumber dependence seems to be diffusive ($q^{-2}$) over a wider range 
than that of the relaxation itself, a further indication that this ``bare
relaxation rate'' is simpler than the full dynamics.
\end{abstract}

\maketitle

\section{Introduction}

Time-dependent correlation functions are a standard tool for quantifying the 
dynamics of physical systems.\cite{Chaikin/Lubensky:1995} For the most basic 
analyses, knowledge of a 
single quantity---the relaxation time---and its dependence on external 
parameters such as temperature, is sufficient. But often more 
detailed information, in particular concerning the shape of the correlation
function, is required. This might be, for example,  because competing 
theoretical explanations
make different predictions about the shape, for example.
In the case of highly viscous glass-forming liquids\cite{Brawer:1985,
Ediger/Angell/Nagel:1996} some functions typically 
investigated include the autocorrelation functions of
energy, pressure, shear-stress, and dielectric moment. These are related 
through the fluctuation-dissipation theorem\cite{Chaikin/Lubensky:1995} 
to the corresponding dynamic 
response functions: frequency dependent specific-heat, bulk modulus, 
shear-modulus (or equivalently viscosity) and dielectric constant, 
respectively. The key features of viscous-liquid dynamics are (1) the 
non-Arrhenius temperature dependence of the main, or alpha,
relaxation time and (2) the
non-exponentiality of the corresponding correlation functions. Any theory of
glass-forming liquids has to be able to explain these ``nons''---violations of 
what is generally expected in relaxing systems (Arrhenius temperature 
dependence and exponential, or Debye, relaxation). 

This paper presents a method for fitting relaxation data in the
time domain, by parameterizing not the correlation itself 
but its
associated memory function.\cite{Boon/Yip:1980} The latter is a unique function
related to the correlation function. The motivation comes from the hope that a
 relatively simple
description of the memory function could be attained, and this hope turns out
to be justified---very good fits can be achieved by taking the memory function
to have the form of an inverse power law for non-zero times. Moreover the 
exponent of the power law seems to have a common value, approaching at low
temperatures a value around 1.6 for the 2D system and 1.5 for the 3D
system, for the different
correlation functions, something that is not true of the ``stretching
parameter'' $\beta$ when fitting using a stretched exponential. In fact 
when the memory function parameterization is generalized to 
include an exponential cut-off, the exponent is temperature independent.

In the next section we discuss
some general ideas for describing relaxation and
review the concept of the memory function associated with a correlation 
function. Following that is a brief description of our simulations, while
Section~\ref{DataAnalysis} presents the details of our analysis method as
applied to time-domain data from the simulations. Results from the fitting
procedure are presented in Section~\ref{Results}.
Section~\ref{Discussion} discusses possible ways of interpreting the power-law
description.

\section{\label{Relaxation}Non-exponential relaxation and the 
memory function}

\subsection{Correlation functions}

In this work we consider normalized correlation functions $\psi(t)$, typically
(but not exclusively) defined in terms of a dynamical variable $A(t),$ as

\begin{equation}
\psi(t) \equiv \frac{\angleb{\Delta A(0)\Delta A(t)}}{\angleb{(\Delta A(0))^2}}.
\end{equation}

\nod For example $A(t)$ could be the potential energy, pressure, or shear 
stress. Here  $\Delta$ indicates deviation from the thermal average.
Debye relaxation corresponds to a simple exponential decay with time constant
$\tau_D$:

\begin{equation}
\psi_D(t) = \exp(-t/\tau_D).
\end{equation}

\nod There are many ways to define a relaxation time $\tau_\alpha$ which agrees 
with $\tau_D$ in the case of Debye relaxation. For our purposes the most 
convenient is the time integral of $\psi(t)$:

\begin{equation}
\tau_\alpha \equiv \int_0^\infty \psi(t) dt
\end{equation}

One advantage of this definition is that it uses all of the information 
contained in $\psi(t)$, unlike, for example, a definition based on the time
at which $\psi(t)$ attains a certain value. It is also convenient when working
with Laplace transforms, since this is simply the the $s=0$ value of the
Laplace transform of $\psi(t)$. The function most commonly used to 
fit time-domain data is the
stretched exponential or Kohlrausch-Watt-Williams (KWW)\cite{Kohlrausch:1854,
Williams/Watts:1970} function,

\begin{equation}\label{KWW}
\psi_{KWW}(t) = \exp\left(-(t/\tau_{KWW})^{\beta_{KWW}}\right),
\end{equation}

\nod where $\tau_{KWW}$ sets the time scale and $\beta_{KWW}$ is a number between
0 and 1, known as the stretching parameter. Exponential behavior is recovered 
for $\beta=1$. By integrating we find that $\tau_\alpha = 
\Gamma(1/\beta_{KWW})\tau_{KWW}/\beta_{KWW}$. Fits to Eq.~(\ref{KWW}) are often 
very good, but on the other hand experiments\cite{Olsen/others:2008} suggest 
that the relaxation crosses over
to exponential at the longest times (a decade or more beyond 
$\tau_{KWW}$), corresponding to the last few percent of relaxation. 
This has not been investigated much,
if at all, in simulations, probably because it is difficult to get high quality
data at the longest times. In the case of single-particle diffusion, however,
the crossover to exponential behavior should coincide with the onset of
 Fickian  diffusion and the convergence of the van Hove correlation function to
 a Gaussian, which have been studied by various 
authors.\cite{Szamel/Flenner:2006b, Stariolo/Fabricius:2006}

\subsection{\label{MemoryFunction}The memory function}

The memory function concept was introduced by the work of
 Zwanzig\cite{Zwanzig:1961} and Mori,\cite{Mori:1965} which provided a 
theoretical formalism for the calculation of correlation functions of
many-body systems, based on projector-operator techniques.
One of the historically most important uses for the memory function has
been to elucidate the short-time structure of correlation functions. For
example, pure exponential decay, represented by a delta-like memory function,
is unphysical at short times, since the correlation, being even and smooth,
should have zero slope at $t=0$. Replacing the delta function with a less 
singular function (for example a step function, a Gaussian, or an exponential)
gives more physical short-time behavior. In this
work, however, we are concerned with the long time behavior, and in our
analysis will ignore the short time behavior, which is mainly associated with
vibrational motion and irrelevant for structural relaxation.

The memory function, $K(t)$, is defined\cite{Boon/Yip:1980} for a normalized 
autocorrelation function $\psi(t)$ by

\begin{equation}\label{memoryFunctionDef}
\frac{d\psi}{dt}(t) = - \int_0^t K(t-t') \psi(t') dt'.
\end{equation}

\nod The relation between $\psi(t)$ and $K(t)$ is completely invertible, which
can be seen by considering the Laplace transformation of this equation:

\begin{equation}
s \tilde \psi(s) - 1 = - \tilde K(s) \tilde\psi(s),
\end{equation}

\nod which implies

\begin{equation}\label{psi_s_K_s}
\tilde\psi(s) = \frac{1}{s+\tilde K(s)},
\end{equation}

\nod an invertible relation between $\tilde\psi(s)$ and $\tilde K(s)$.
Thus we can think of the memory function as a kind of transform of the 
correlation function. Setting $s=0$ in Eq.~(\ref{psi_s_K_s}) gives 
an expression for $\tau_\alpha$:

\begin{equation}\label{tau_alpha_from_K}
\tau_\alpha = \frac{1}{\int_0^\infty K(t)dt}.
\end{equation}

This will be useful below. Although though of our analysis will involve a pure 
power-law decay of $K(t)$, it should be noted
that if $\psi(t)$ is known to decay exponentially at the 
longest times, the same must be true of the memory function: If 
$\psi(t)$ decays as an exponentially $\sim \exp(-\lambda t)$ at long times 
(with $\lambda$ is real and positive), this implies that the least negative 
singularity of $\tilde \psi(s)$ is at $s=-\lambda$. In particular $
\tilde \psi(s)$ is analytic,
as well as real and positive, in a finite region about $s=0$ 
(recall $\tilde \psi(s=0)=\tau_\alpha$).
The same is true of $1/\tilde\psi(s)$ and therefore $\tilde K(s)$ 
(=$1/\tilde\psi(s)-s$). Being finite and analytic at $s=0$ is incompatible
with a power-law decay for $K(t)$, which would correspond to a
power-law for $\tilde K(s)$ with (in general) non-integer exponent as 
$s\rightarrow0$. If an exponential cut-off is included in $K(t)$, however, this
 moves the singular point
a finite distance to the left along the negative real $s$-axis.

\subsection{Interpretation as a rate}

A useful starting point for
considering the meaning of $K(t)$ is the case $K(t)=\gamma\delta(t)$, which 
gives exponential 
relaxation $\psi(t)=\exp(-\gamma t)$, i.e., $\tau_\alpha=1/\gamma$, consistent
with Eq.~(\ref{tau_alpha_from_K}). This is the memory-less, or Markov, case.
Note that the `amplitude'  $\gamma$ gives the rate of the exponential decay. 
Consider integrating Eq.~(\ref{memoryFunctionDef}) as if
it were a physical equation of motion: In the memoryless case, as time goes on,
the delta function means that the only contribution to the integral on the
right-hand side is from the current time, $t'=t$. Thus, the rate of change of 
$\psi(t)$ is proportional to its current value and 
independent of its previous values. This of course gives exponential 
relaxation. If we generalize slightly to the case $K(t)=K$, a constant,
for $t<T$, and zero otherwise, then for times much greater than $T$ we again
expect exponential decay with relaxation time 
$\tau_\alpha=1/(KT)$.\footnote{If $T$ is
sufficiently large for a given value of $KT$, 
there will also be an oscillatory factor.} When $K(t)$ is non-zero 
for all $t>0$, but $\int_0^t K(t')dt'$ converges sufficiently quickly (e.g. 
exponentially) to a finite positive value, we can expect exponential decay 
in $\psi(t)$ for long times.

For times $t$ when the memory function is still changing, $\int_0^t K(t')dt'$
gives an estimate of the instantaneous decay rate of $\psi(t)$.
What happens when $K(t)$, which must be positive
at time zero (otherwise $\psi(t)$ will diverge), becomes negative? At that 
point $\int_0^tK(t')dt'$ starts to decrease, implying that the decay rate 
decreases. As long as $K(t)<0$, $\psi(t)$ will exhibit
 slower-than-exponential relaxation---the instantaneous decay rate keeps
decreasing. Therefore we expect that for viscous liquids
 the memory function is negative and significant for times up to $\tau_\alpha$,
around which time $\int_0^t K(t')dt'$ has more or less converged to the limiting
value and exponential relaxation takes over.

\subsection{Discrete memory function}

When we consider simulation data we will be not interested in short-time 
vibrational contributions to $\psi(t)$ or $K(t)$. A standard way of removing
the effects of vibrations is to consider the so-called inherent 
dynamics,\cite{Stillinger:1995} obtained by minimizing configurations along the
actual simulated trajectory to the corresponding local minimum of the
potential energy---the inherent state. The correspondence is defined by 
steepest-descent minimization and is, in principle, unique for almost all 
configurations. It has been shown that the resulting trajectory of inherent 
states preserves the long time features of the 
dynamics. \cite{Schroder/others:2000,
 Liao/Chen:2001} Correlation functions are essentially 
unchanged except that the usual initial decay due to vibrations on the
picosecond time scale is missing. A technical problem is introduced by this
procedure, however: The inherent correlation functions tend to have non-zero,
and therefore discontinuous, slopes at $t=0$. This is not so surprising, given 
that the inherent trajectory is itself 
necessarily discontinuous, but it does pose
problems for defining the short-time behavior of the memory function. To avoid
these problems, and given that simulation data is generally available at 
regular, discrete times, we write a discrete version
of Eq.~(\ref{memoryFunctionDef}):

\begin{equation}\label{memoryFunctionDiscrete}
\psi_{n+1} - \psi_n = - \sum_{m=0}^n K_{n-m} \psi_m.
\end{equation}

Here time has been discretized in units of
$\Delta t$, so $\psi_n\equiv\psi(t=n\Delta t)$. 
Equation~(\ref{memoryFunctionDiscrete}) can be straightforwardly
used to calculate the discrete
values of the autocorrelation function, $\{\psi_n\}, n\geq 0$, 
given the discrete values of the memory function, $\{K_m\}, m \geq 0$: 
Start by setting $\psi_0=1$ and then for each $n>0$ in turn,
calculate the sum on the right hand side, which only involves $\psi_m$
for $m \leq n$ and the known $\{K_m\}$, to get the next unknown
value, $\psi_{n+1}$. It is almost as straightforward to see that the reverse
transformation is possible, i.e., given $\{\psi_n\}, n\geq 0$, to determine
$\{K_m\},m \geq 0$. For example, from Eq.~(\ref{memoryFunctionDiscrete})
for $n=0$ one finds

\begin{equation}
K_0=1-\psi_1/\psi_0 = 1-\psi_1,
\end{equation}

\nod (since we assume
the normalization $\psi_0=1$). Knowing now $K_0$, writing 
Eq.~(\ref{memoryFunctionDiscrete}) for $n=1$ gives again an equation with
only one unknown, namely $K_1$. In this way the values of $K_m$ can be solved
for one by one. This process may be formalized using the idea of a 
generating function.\cite{Wilf:1994} We consider the
series $\{\psi_n\}$  as the coefficients of a
power series in a variable $z$, which (formally) defines the
so-called generating function $\psi(z) \equiv \sum_0^\infty \psi_m z^m$, 
similarly $K(z)\equiv \sum_0^\infty K_m z^m$. We may
treat generating functions as quantities which can be added, multiplied
and divided (providing the first coefficient is non-zero), and thus can note
that the right hand of Eq.~(\ref{memoryFunctionDiscrete}) is in fact the $n$th
coefficient in the product series $-K(z)\psi(z)$ (this is like  the
convolution-product correspondence in Laplace transforms).
Thus the equation may be re-written as

\begin{equation}
(\Delta\psi)(z) = - K(z)\psi(z),
\end{equation}

\nod where the difference function $(\Delta\psi)(z)$ is defined as the series
$(\Delta\psi)_0=0$, $(\Delta\psi)_n=\psi_n-\psi_{n-1}, n>0$. $K(z)$ is then
given by 

\begin{equation}
K(z) = - \left( \frac{1}{\psi(z)}\right)(\Delta\psi)(z),
\end{equation}

\nod which may be evaluated using standard algorithms for multiplying and
dividing power series.\cite{Wilf:1994} Mathematically this procedure is
equivalent to the inversion procedure mentioned above; thinking in terms of
generating functions simply makes the implementation more straightforward.
We use this transformation in the data analysis to relate correlation functions
to their corresponding memory functions. In the following, we use $K(t)$ to 
mean $K_m/(\Delta t)^2$, where $t=m\Delta t$.

\section{\label{Simulations}Simulations}

\subsection{Systems and potentials}

We have simulated both two- and three-dimensional (2D and 3D) binary
Lennard-Jones (BLJ) fluids. The parameters for the BLJ potential for both
2D and 3D are 
(where L and S stand for large
and small particles, and $\epsilon$ and $\sigma$ the
energy and length scales, respectively) $\epsilon_{LL}=1$, $\epsilon_{SS}=0.5$, 
$\epsilon_{LS}=1.5$, $\sigma_{LL}=1$, $\sigma_{SS}=0.88$,
$\sigma_{LS}=0.8$. All particles have the same mass $m=1$. These
parameters are identical to those of the 3D BLJ
introduced by Kob and Andersen \cite{Kob/Andersen:1994}. The potential was
truncated using an interpolating polynomial between 2.4 $\sigma_{\alpha\beta}$ 
and  2.7 $\sigma_{\alpha\beta}$ ($\alpha,\beta \in \{L,S\}$). 
In all simulations periodic boundary conditions with constant area/volume
were used. In 2D the total number of particles  $N_p$=700 of which 60\% were
of type L, while in 3D, $N_p$=1372 of which 80\% were of type L. The particle
density was 1.2$\sigma_{LL}^{-d}$ in both cases (where $d$ is the dimension).
Constant energy dynamics were simulated using the Verlet algorithm with a time
step of 0.01 $\sigma_{LL}\sqrt{m/\epsilon_{LL}}$
From now on, all quantities
 will be reported in the units defined by $\epsilon_{LL}$,
 $\sigma_{LL}$ and $m$. If one converts to physical units by assuming that 
the parameters for large particles are chosen to model Argon
(i.e., $\sigma_{LL}=0.34$ nm, $\epsilon_{LL}=997$ kJ/mol, $m=39.95$ u), then the
time unit corresponds roughly to 2 ps.

\subsection{Runs}

Initial configurations consisted of fcc lattices
with particles randomly assigned to be of type L or S such that the appropriate
fraction was achieved. All simulations were at constant NVE; in particular the
energy was controlled rather than the temperature $T$,
 but we quote the temperature
defined in terms of the mean kinetic energy per particle instead as it is more
meaningful. Starting at at $T\sim 1.0$ (where the
 liquid is not particularly viscous) ten independent runs (differing by the
random placement of the particles on the initial lattice) were made. The
initial lattices melted immediately and equilibrated in the liquid state.
For each state point, an equilibration run was carried out followed by a 
production run of the same length. Equilibration was checked by examining the 
self-intermediate scattering function $F(q_m,t)$ for the larger particles;
$q_m$ is 6.5 in 2D and 7.3 in 3D (corresponding roughly to
the peak of the structure factor for large particles). $F(q_m,t)$ was required
to be essentially the same for all ten runs, and the run-length to be of order
100 times the structural relaxation time ($\tau_\alpha\equiv \int_0^\infty 
F(q_m,t)dt$). If the condition was not satisfied, the
production run was considered to be part of the equilibration and a new 
production run was initiated from the final configurations.
Steps in temperature were made by changing the total energy according to the 
estimated specific heat; a new kinetic energy was chosen such that when summed
with the current potential energy the correct total energy would result. Then
the particles' velocities were randomized using a Gaussian distribution with
the appropriate width and then rescaling to give the exact desired kinetic
 energy. During the equilibration process it was checked that the new 
temperature (mean kinetic energy) was the desired one and velocities were 
rescaled if necessary (generally a small change).

\subsection{Calculating correlation functions}

During the simulations, collective quantities such as potential and kinetic
energy and the different components of stress were computed and written at
regular intervals (typically every 100 or 500 time steps, though some shorter
runs were made with more frequent output in order to determine the short time
behavior of the correlation functions).
Configurations were saved at ``logarithmic intervals'' to allow calculation of 
$F(q,t)$ for a broad range of times $t$ without writing configurations as
 frequently as the collective quantities (which would require a
 huge amount of disk storage). Because $F(q,t)$
involves an average over particles, it has significantly smaller statistical
error and therefore has low noise out to quite long times beyond the relaxation
time. It should be noted, however, that it is technically not the 
autocorrelation function of a dynamical variable.
Autocorrelation functions of collective quantities were calculated using 
standard Fourier transform techniques. Both these and $F(q,t)$
were averaged over the ten runs. For the main part of the analysis the 
autocorrelation functions were also
 ``logarithmically averaged'' in time. This weights 
different time scales equally and involves grouping the data points into 
equal-sized bins of $\log(t)$, and averaging within each bin.

\subsection{Inherent states}

Some of the analysis was carried on inherent-state configurations. The method
used to minimize the energy was a combination of ``molecular dynamics
 minimization''\cite{Stoltze:1997} and conjugate 
gradient.\cite{Press/others:1987} This combination has been found to reduce
the possibility of finding the ``wrong'' minimum, which can happen if 
conjugate gradient alone is used.\cite{Bailey/Schroder/Dyre:2009}
Configurations were minimized every 10, 100 or 1000 configurations---more 
frequent minimization enables resolution of shorter time scales, but longer 
time scales cannot be accurately probed.

\section{\label{DataAnalysis}Data analysis: Determining $K(t)$}

\begin{figure}
\epsfig{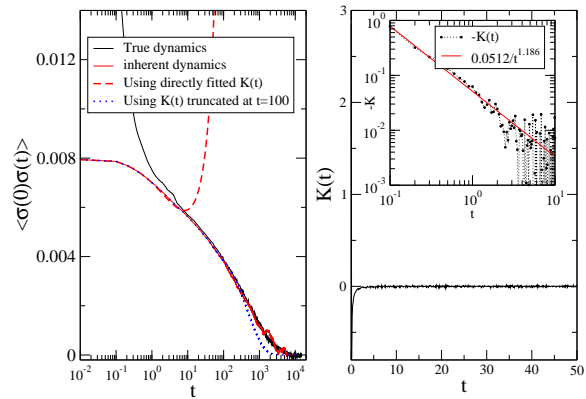}
\caption{\label{shearCorrMem2D0.34} (Color online)
Left panel, shear stress (unnormalized)
autocorrelation function for true and inherent dynamics. They differ only
at short times. The slight dip in the curve for true dynamics at $t=5$ is 
from combining separate short- and long- time simulations. 
Minimization 
was carried out every ten time steps, which defines the resolution of
 the memory function, $\Delta t=10 dt=0.1$.
The dashed line shows the correlation function obtained from attempting
to invert the attempted fit to the memory function, while the dotted line
 shows that obtained by inverting using the exact memory
function truncated at $t=100$. Right panel, memory function for
the inherent shear-stress autocorrelation function. Inset shows the negative 
tail in a double-logarithmic plot, along with a power-law fit.} 

\end{figure}

\subsection{Inversion of the correlation function to get the exact $K(t)$}

We start by applying the transformation described above to the simulation data.
Fig.~\ref{shearCorrMem2D0.34} shows the 2D, T=0.34 shear-stress 
autocorrelation function 
for both true and inherent dynamics. The agreement between
the two at long times is clear, although there is more noise in the inherent 
case---due to the cost of minimization, not as long times can be simulated.
The right
panel of Fig.~\ref{shearCorrMem2D0.34} shows the result of the discrete 
memory transform. Recall that this returns the exact memory function $K(t)$ for
the data. This is positive at $t=0$, but negative for all $t>0$ (note that 
$t$ really means $n\Delta t$ for an integer $n$; since minimization was carried
out every ten time steps,  $\Delta t=0.1$). As shown in 
the inset, the data suggest an inverse power-law for the negative tail,
 although the noise is too great to justify 
this for $t>2$, a very short time compared to the relaxation time 
$\tau_\alpha \sim 3000$. Nevertheless let us attempt a fit to a 
simple power-law in the hope that this is in fact the correct form of $K(t)$
for $t>0$. The fit yields the expression $0.0512/t^{1.186}$ for $K(t)$.
Testing whether this expression is in fact a good representation of the true
memory function is simple: we simply use Eq.~(\ref{memoryFunctionDiscrete})
to construct the correlation function corresponding to the fitted $K(t)$. 
The value of $K_0$ is not fit; rather the exact value from the inversion
is used. The result, shown as the dashed line in the left panel of 
Fig.~\ref{shearCorrMem2D0.34}, is a disaster: after reasonable 
agreement up to $t\sim 10$, the transformed $\psi(t)$ stops decaying and in
fact begins to increase and eventually diverges. The problem can actually be
foreseen by considering the time integral of $K(t)$, which in the discrete
case is the sum $ \sum_n K_n / \Delta t$. 
The value of $K_0 (=1-\psi_1)$ here is 0.0286, so that its contribution
to the sum is 0.286. The sum over the power-law part requires the evaluation
of the Riemann zeta function $\zeta(s)$ at $s=1.186$, which is estimated
numerically to be 5.97. Including the appropriate factors of $\Delta t$ and
the prefactor yields the contribution from the power law to be -0.469, 
making the whole sum negative. Moreover the
sum first becomes negative around where the inverted correlation function
begins to increase--at this point the effective decay rate has become negative,
and so the correlation function grows rather than decays.

Clearly, direct fitting of the memory function is problematic. The fit which 
applies at least up to $t\sim 2$ clearly is wrong for $t>10$. Because the
noise in the exact memory function is so much larger than the true value, 
standard fitting procedures cannot distinguish it from zero in this region. On
the other hand, we can directly check that it is indeed significantly different
from zero for quite long times: Taking the exact memory function, but setting
it to zero for times greater than $t=100$, and inverting, we find the 
correlation function indicated by the dotted line in 
Fig.~\ref{shearCorrMem2D0.34}. After $t\sim 500$ it drops below the true
correlation function in a more or less exponential fashion, as
expected when the memory function is only non-zero for a finite range of time.
Thus the apparent power-law behavior at short times must cross over to
something more rapidly decaying, but still significantly different from 
zero.

\begin{figure}
\epsfig{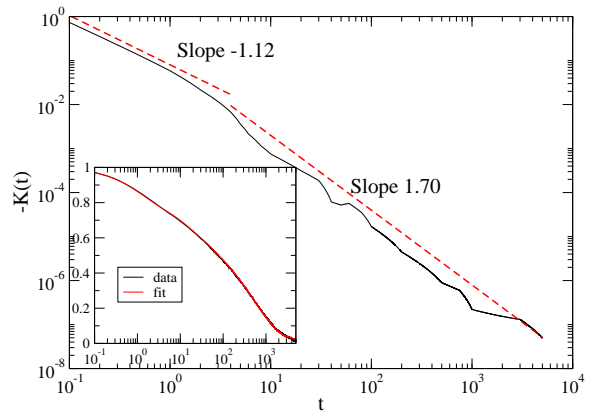}
\caption{\label{fittedAllTime2D_T0.34}Main figure, fitted memory function
 consisting of an inverse power-law multiplied by a piecewise linear 
function. Inset, comparison of the shear stress autocorrelation function from 
simulations and the correlation function corresponding to the fitted memory
function. In this case the correlation function was a combination of that
determined by inherent dynamics at short time and that determined by real
dynamics at long-time, in order to have a more accurate and less noisy curve at
long times.}

\end{figure}

\subsection{\label{MatchingCorr}Matching the correlation function by fitting $K(t)$.}

An approach other than direct fitting of the exact $K(t)$ is clearly needed.
Since the noise in the correlation function is small compared to the function
itself, comparing this to a trial function in a fitting procedure
will not suffer the same problem.
This suggests a practical method for determining the memory function: We choose
a functional form which is quite flexible, containing many parameters. For
any given set of parameter values we can generate the corresponding trial
correlation function which can be compared in a least-squares sense to the 
actual correlation function. Then we vary the parameters to optimize the 
match.

The value of $K_0$ is fixed from the start to $1-\psi_1$.
Given that the exact $K(t)$ seems to have an initial inverse 
power-law decay for $t>0$, we start with such a function.
To provide flexibility we multiply by a piecewise linear function; such a 
function will have discontinuous changes in slope but is simple and 
general---and hopefully the dominant behavior has been captured by the 
power-law factor. Thus this general fitting form $K_{gen}(t)$ is

\begin{equation}
K_{gen}(t) = -\frac{A}{t^\alpha}L(t), t>0
\end{equation}

\nod where $L(t)$ is a piecewise linear function defined by a set of nodes 
$t_k$ and values $L_k$ for $k=0,1,2,\ldots,n_{nodes}$. The first node is at 
$t=0$, i.e., $t_0=0$ and $L_0$ is fixed at unity. We choose $n_{nodes}\sim20$ to
provide sufficient flexibility; a typical set of
node locations $t_k$ is 0, 1, 2, 4, 5, 6, 8, 10, 20, 30, 40, 50, 60, 80, 100,
200, 500, 750, 1000, 3000, 5000. The adjustable parameters are then the overall
coefficient $A$, the exponent $\alpha$, and the coefficients
 $L_k$, $k=1,\ldots, n_{nodes}$. At this
stage we do not seek the true functional form of $K(t)$; rather, we wish to
have a fairly accurate representation which is relatively free of noise. To 
obtain the optimal parameter values in $K_{gen}(t)$, a conjugate-gradient
procedure is used. Some details of this are given in the appendix.
Fig.~\ref{fittedAllTime2D_T0.34} shows the result of this procedure. While the
discreteness due to the piecewise linear function is clearly visible, the
overall form is clear. In particular, there is
a clear cross-over at around $t=5$ from the power law we initially identified
to a more rapidly decaying power law,
with exponent $\sim$1.7. This cross-over means that the time integral converges
to a positive value and the effective decay rate never becomes negative.

\begin{figure}
\epsfig{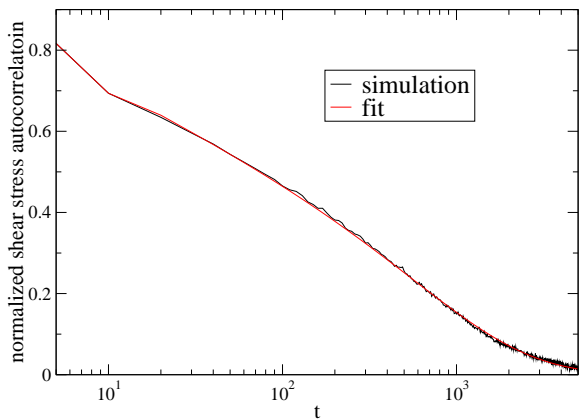}
\caption{\label{singlePLfit2D_T0.34}.  (Color online)
Comparison of autocorrelation function
and fit with a memory function involving a single inverse power law in the
tail. The data is the same as in Fig.~\ref{fittedAllTime2D_T0.34}, but only 
times which were multiples of 10 were included. The fitted exponent was 1.84, 
reasonably close to the apparent exponent of 1.7 observed in 
Fig.~\ref{fittedAllTime2D_T0.34}.}
\end{figure}

\subsection{Simple power-law form for $K(t)$ for $t>10$.}

The observation that $K(t)$ is well approximated by a single 
inverse power law for times greater than about 5
immediately suggests a simplification: If we choose
to not resolve times shorter than this, a fit using only a single power
law for the non-zero part of $K(t)$ may be possible. 
Fig.~\ref{singlePLfit2D_T0.34} shows that this is indeed true. Here times 
greater than 10 were included, corresponding to sampling at an interval
 $\Delta t = 10$. This time scale 
is also significant for another reason (not necessarily unrelated): It
corresponds to the end of the vibrational contribution to the stress 
autocorrelation function of the true dynamics (see left 
panel of Fig.~\ref{shearCorrMem2D0.34}). This suggests that we can avoid the 
time consuming energy minimization process altogether and use only the data 
from the true dynamics. There is one problem, though: We cannot get the $t=0$ 
value of the (inherent) autocorrelation function from the true dynamics. To get
 around this we leave it as a fitting parameter; this 
gives us a procedure for fitting 
autocorrelation functions with three parameters: $R_{01}\equiv\psi_0/\psi_1$, 
$A$ and $\alpha$.
Note that $R_{01}$ is directly related to $K_0$,
so one could equivalently think of $K_0$ as the parameter. This fitting
function involving a single inverse power law
has been used in most of the analysis.

It is convenient to represent that amplitude of the power-law in terms of a 
``tail-weight fraction'' $f$, defined such that $f=1$ corresponds to the limiting case
where the sum over the whole negative ($m>0$) part of $K_m$ 
exactly cancels the positive contribution from $K_0$. Thus we represent $K_m$
for $m>0$ as

\begin{equation}\label{power_law_Km}
K_m = - K_0 \frac{f}{\zeta(\alpha)}\frac{1}{m^\alpha}, m>0, 
\end{equation}

\nod where $\zeta(\alpha)\equiv \sum_{n=1}^\infty 1/n^\alpha$ is the Riemann 
zeta-function. Then the relaxation time is given simply by

\begin{equation}
\tau = \frac{1}{\sum_n (K_n/(\Delta t)^2)\Delta t} 
 = \frac{1}{K_0 \Delta t} \frac{1}{1-f}
\end{equation}

This form makes explicit a few things: for near zero $f$, the relaxation rate 
is essentially the inverse of $K_0$. As $f$ increases the negative tail 
``cancels out'' part of the latter, increasing the relaxation time, and the 
limit $f\rightarrow 1$ gives a diverging relaxation time. 

\begin{figure}
\epsfig{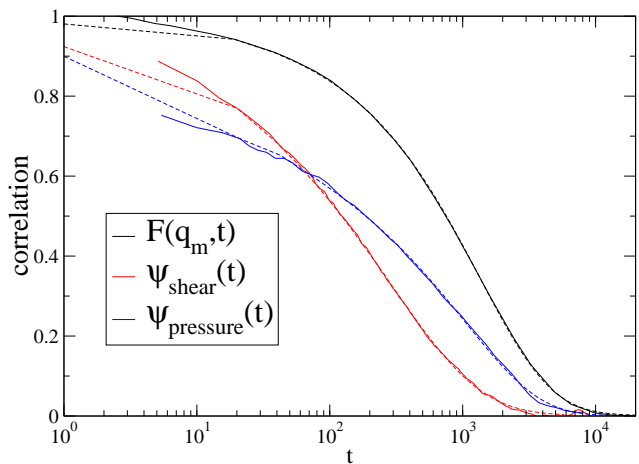}
\caption{\label{demonstrateFitsISFShearPress} (Color online)
Main figure, examples of fits 
using the 3-parameter power-law form of $K(t)$ to three correlation functions
calculated from simulations of the Kob-Andersen binary Lennard-Jones mixture
in 3D at $T$=0.46. Solid lines: data; dashed lines: fits. In contrast to
 Fig.~\ref{singlePLfit2D_T0.34}, logarithmic binning/averaging was used for
the shear stress and pressure autocorrelation functions.}
\end{figure}

Examples of fits to the three functions, $F(q_m,t)$ and the shear stress 
and pressure autocorrelation functions, are shown in the main part of 
Fig.~\ref{demonstrateFitsISFShearPress} for one temperature in the
3D system. Here and in the all the analysis presented below, the discretization
interval was $\Delta t$=20.

\subsection{Long-time behavior: An exponential cut-off in $K(t)$}

\begin{figure}
\epsfig{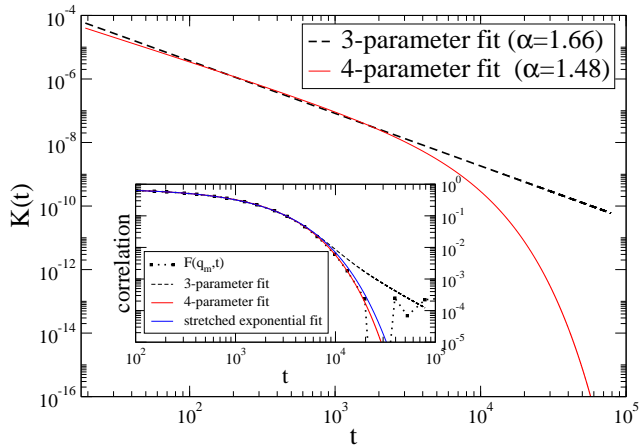}
\caption{\label{compareISF_3D_0p46withK} (Color online)
Comparison of memory functions obtained for intermediate scattering function
(3D, $T=0.46$, $\tau_\alpha=1500$) using 3- and 4-parameter fits. The exponential cut-off in the
latter is clearly visible.
Inset, comparison of the correlation functions from the fit with the data, 
along with a stretched exponential. Here a
double-log scale is used to emphasize the long-time behavior (the differences
would not be visible otherwise).}
\end{figure}

As mentioned in Section~\ref{Relaxation}, experiments\cite{Olsen/others:2008} 
suggest that autocorrelation functions switch over to
exponential decay at very long times. This is incompatible with the fitting
function we have identified---as discussed in Section~\ref{Relaxation}, 
power-law decay of $K(t)$ corresponds to a power-law singularity in $\tilde 
K(s)$ as $s\rightarrow0$, implying non-analyticity of $\tilde\psi(s)$
in the same limit, which excludes exponential behavior of $K(t)$ at long times.
Thus we should---in principle at least---allow the possibility of including
an exponential cut-off in the fitting function for $K(t)$. The question is
whether the available data requires it to get a satisfactory fit.
Only in the case of the intermediate scattering function is the noise 
sufficiently low to allow investigation of the long time behavior (when the 
correlation function has decayed to less than 1\%  of its original value).
Fig.~\ref{compareISF_3D_0p46withK} compares the three-parameter
fit using Eq.~\ref{power_law_Km} with a 4-parameter fit involving an 
exponential cut-off:

\begin{equation}\label{exp_power_lawKm}
K_m = - K_0 \frac{f}{\zeta(\alpha)}\frac{\exp(-m \Delta t/\tau_c)}{m^\alpha}, m>0
\end{equation}

\nod where $\tau_c$ is the characteristic time of the exponential cut-off.
The memory functions are shown in the main part of the figure, while the 
correlation functions are shown in the inset.
For comparison, a stretched-exponential fit is also 
included. The three-parameter (pure power-law) 
memory-function fit noticeably underestimates the decay rate at long times.
Including the cut-off gives a much better fit, although some improvement is of 
course expected due to the extra parameter). Notice that
the stretched exponential function fits also better than the pure power-law fit
at long times, as well as the cut-off memory 
function. For the shear stress and pressure autocorrelation
functions the long time data is not good enough to warrant including the 
exponential cut-off as a fourth parameter, although we shall see that it can 
make sense to do so when there is reason to fix the exponent $\alpha$ to a 
particular value. In the following, ``four-parameter'' and ``three-parameter'' 
fits refer respectively to whether the exponential cut-off was included or not,
regardless of whether one or more of those parameters was constrained in the 
fitting process.



\section{\label{Results}Results}

\subsection{Relaxation times}

\begin{figure}
\epsfig{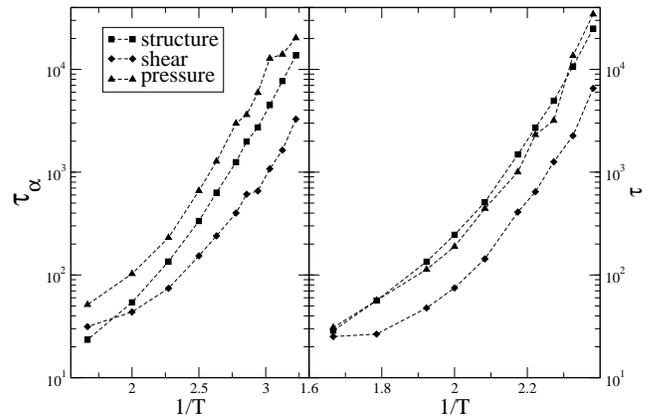}
\caption{\label{showAllRelaxationTimes} Arrhenius plot of relaxation times
for structure ($F(q_m, t)$), shear stress and pressure autocorrelation
functions, for 2D and 3D data (closed symbol), determined by summing the
fitted normalized memory function.}
\end{figure}

We start by presenting the relaxation times $\tau_\alpha$ for the different 
correlation functions. As stated earlier, we define $\tau_\alpha$ as the 
integral of the normalized correlation function. With simulation data there is
a contribution at short times due to vibrations. The relative height of this
varies among the different correlation functions, making the choice of
normalization problematic. In principle this can be removed by considering
correlation functions of inherent quantities, but the time taken for quenches 
means that determining the correlation function accurately at long times is 
difficult. We avoid these problems by using the fitted correlation functions,
normalizing by the zero-time value that emerges from the fitting process. 
To be consistent with the discrete-time formalism for dealing with the memory 
function we sum the correlation functions and multiply by $\Delta t$,
rather than integrating them. This makes a positive difference of order 
$\Delta t/2$, negligible except when the relaxation
time is of order $\Delta t$. This is the case at the highest temperatures, 
particularly for shear stress (note the upwards bend
in the relaxation time curve at the extreme left of the plot).
The value of $\tau_\alpha$ determined from the fit varies only
slightly according to whether the 3- or 4- parameter fitting function for
$K(t)$ is used. 
Fig.~\ref{showAllRelaxationTimes} shows Arrhenius plots for both 2D and 3D
 systems of the three correlation functions investigated. Some 
curvature---corresponding to so-called ``fragility'' of the
viscous liquid---is evident, although less so in the 2D data. 
In both 2D and 3D the shear relaxation time is noticeably
shorter than those of pressure and structural relaxation ($F(q_m,t)$). The
pressure relaxation time tracks closely the structural one in 3D, but exceeds 
it noticeably in 2D. 
In fact the difference between pressure and shear stress relaxation
in 2D is of order a factor of ten; in 3D it is closer to a factor of four.
Note that at higher temperatures (0.5--0.6) the relaxation time is of the same 
order as the discretization interval $\Delta t=20$, so a large part of the
relaxation actually takes place within the first interval. Unsurprisingly, this
limits the ability of the fitting procedure to accurately determine the 
zero-time value which leads to errors in normalization and hence in
$\tau_\alpha$, particularly for the collective correlation functions of pressure
and shear stress where the noise is high. In these cases a fit to a stretched
exponential was made first, which was then used for the memory function fit.
We note finally that is apparently nothing special happening around the
mode-coupling temperature $T_c=0.435$  previously identified 
for the 3D Kob-Andersen system (albeit with a slightly different cut-off in the
potential).

\subsection{Short-time rate $K_0$ and corresponding time $\tau_s$}

\begin{figure}
\epsfig{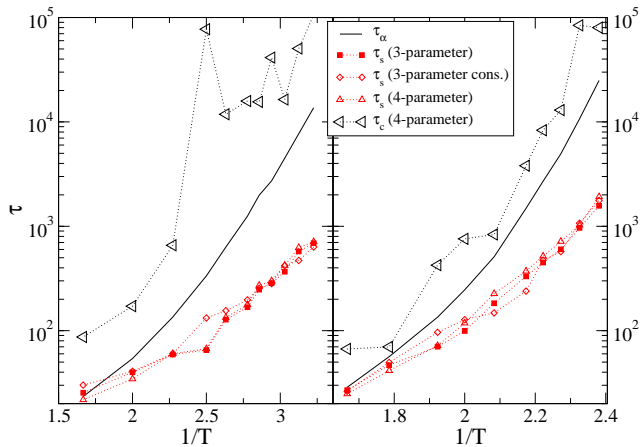}
\caption{\label{ISF_timeParameters} Arrhenius plot of the inverse short-time
rate $\tau_s = \Delta t/K_0$ determined from different fits to $F(q_m,t)$, 
compared with the relaxation time $\tau_\alpha$, and the cut-off time $\tau_s$ of
 the 4-parameter fit,
for 2D (left panel) and 3D (right panel) systems. The data labelled 
``3-parameter (cons.)'' refers to simultaneous fits of the three correlation
functions at a given temperature, constraining the exponent $\alpha$ to be
the same in all three.}
\end{figure}

\begin{figure}
\epsfig{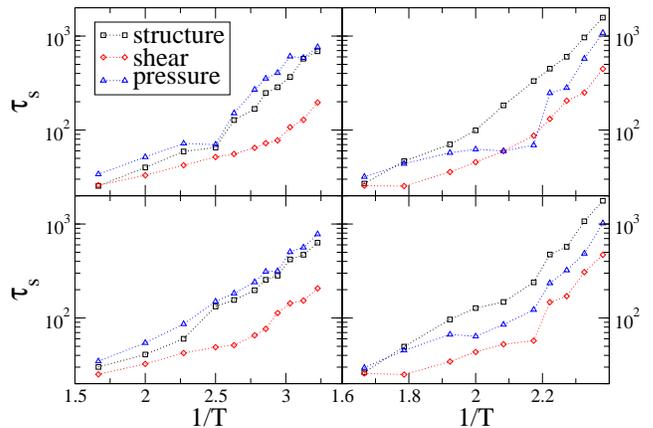}
\caption{\label{IndivMultiK0Tdep}  (Color online)
Comparison of inverse short-time rate 
$\tau_s=\Delta t/K_0$ for different kinds of correlation functions. Left and 
right
 panels show 2D and 3D data, respectively. Upper panels show the results of
fitting each curve separately; lower panels show the results of constraining
the power-law exponent $\alpha$ to be the same for all three curves at a given
temperature.}
\end{figure}

We now investigate the parameters determined by
the fitting procedure starting with the
temperature dependence of the parameter $K_0$, related to the short-time
relaxation rate. More precisely, we consider $\tau_s\equiv\Delta t/K_0$, 
which is an 
effective time scale quantifying the amount of relaxation that occurs over the
interval $\Delta t$, a ``short-time relaxation time''. 
This is plotted Figure~\ref{ISF_timeParameters} along
with $\tau_\alpha$ for $F(q_m,t)$. Results of three different fitting schemes
are shown: three-parameter fit, four-parameter fit (i.e., including the 
exponential cut-off) and a three-parameter fit 
where the exponent $\alpha$ is constrained to
be equal for the three correlation functions at a given temperature. The 
motivation for the latter will become clear later on. The variation between the
different fits gives an estimation of the error bars on this quantity. $\tau_s$
 is less than $\tau_\alpha$, which is necessarily the case
if $K_m<0$ for $m>0$. The main point is
that $\tau_\alpha$ is more or less equal to the $\tau_s$ at high 
temperatures---something necessarily true for exponential relaxation---but 
increases relative to it as temperature decreases. The temperature dependence
of the $\tau_s$ is activated, which is to say it is at least Arrhenius.
The data in the figure are not precise enough to determine whether the rate
is super-Arrhenius, but certainly the effective barrier (the slope in the
figure) is lower than for $\tau_\alpha$ itself.
Figure~\ref{IndivMultiK0Tdep} shows comparison of $\tau_s$ for the three 
different quantities. The upper panels represent independent fits, while the
lower ones represent constrained fits where $\alpha$ is the same for the
three quantities.

\subsection{Power-law exponent $\alpha$}

\begin{figure}
\epsfig{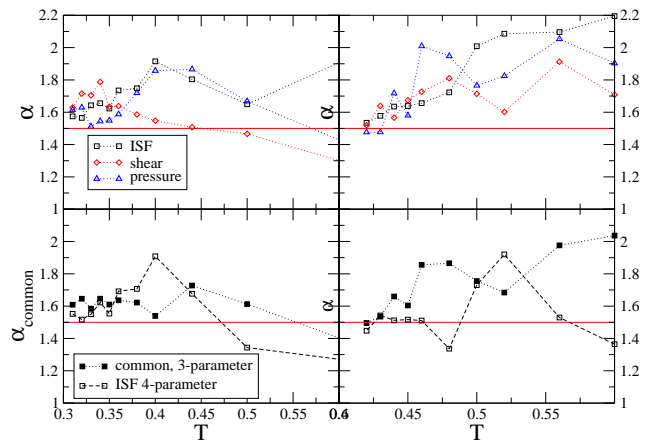}
\caption{\label{showAllAlpha} (Color online)
Temperature dependence of the exponent $\alpha$ 
for different quantities in 2D (left panels) and 3D (right panels) simulations.
Upper panels show the result of independently fitting the different 
correlation functions; lower panels show the common exponent obtained by 
constraining it to be the same for all three, and that obtained in the 
(unconstrained) four-parameter fits to $F(q_m,t)$. In both systems, $T=$0.6 
values are from fits using $\Delta t=$10.
The horizontal lines indicate $\alpha=$1.5.}
\end{figure}

Figure~\ref{showAllAlpha} shows the temperature dependence of the power-law
exponent $\alpha$ for the two different systems, the three different 
correlation functions, and for different fitting schemes. Considering first the
top two panels, the interesting feature is that the exponent values seem to
converge as the temperature is lowered, towards $\sim$1.6 in 2D and $\sim$1.5 
in 3D. At higher temperatures there is significant scatter, and, for the 3D 
case at least,
generally higher values up to around 2. The apparent convergence at low 
temperatures suggests that perhaps the true values of the exponents are
in fact equal for the three functions at each temperature; 
the scatter in the unconstrained fits would then be understood to be due to
noise in the data. We therefore attempted a fit where $\alpha$ is
constrained to be the same for all three correlation functions. In most cases
the quality of the fit is indistinguishable by eye from the unconstrained 
fit (not shown).
The exceptions are some fits to $F(q_m,t)$ in 2D where the constrained fit tends
to reduce the value of $\alpha$ compared to the unconstrained fit, which
results in a somewhat slower decay at long times (this could be probably
compensated for by including the exponential cut-off, as
explained in the following paragraph; we have not done this).
The temperature dependence of the constrained-$\alpha$ is more or less similar
to the constrained case for the 3D data, while noticeably reduced in the 2D
case. The effect on the constrained fit on the values of $\tau_s$ can be seen in
the lower panels of Fig.~\ref{IndivMultiK0Tdep}. There is a tendency towards
smoother temperature dependence, which supports the idea that a constrained fit
can yield more accurate results because the tendency to fit noise in any one
of the curves is limited by the constraint. On the other hand the temperature
dependence of $\tau_s$ for the intermediate scattering function is actually
less smooth in the constrained fit it has been ``infected'' by the
greater noise in the collective functions. In principle this could be 
compensated for assigning a greater weight to the intermediate scattering
function due to the smaller noise. This has not been attempted; it would 
require an unbiased estimate of the errors on the correlation functions.

Also shown in the lower panels of Fig.~\ref{showAllAlpha} are values of
$\alpha$ for unconstrained 4-parameter fits to $F(q_m,t)$. What is noteworthy
here is that while values at higher temperatures still show appreciable 
scatter, there seems to be faster convergence at low temperatures, to values
close to 1.55 and 1.50 in 2D and 3D respectively
 (see in particular the four lowest
temperature points in the 3D case). This suggests the interesting possibility
that the exponent could be in fact be independent of temperature as well as of
which function is considered, and independent of which of the two systems. The
apparently higher values at higher temperatures could be due to not including
the exponential cut-off, which, like a higher value of $\alpha$, would induce
faster relaxation than a pure $\alpha=1.5$ power law. An attempt to find a
temperature-independent $\alpha$ will be described below; first we consider
the results for the parameter $f$.

\begin{figure}
\epsfig{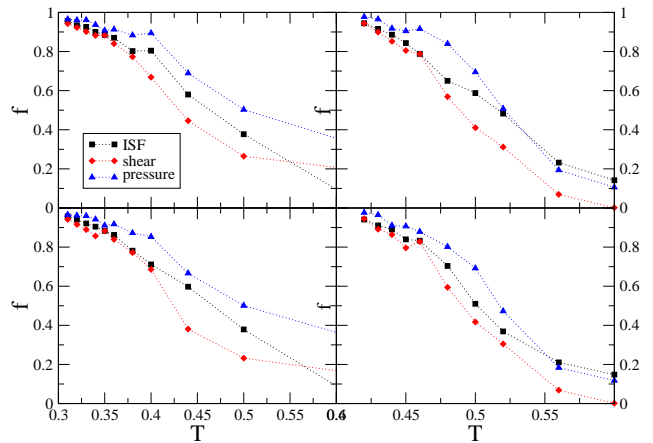}
\caption{\label{showAllF} (Color online)
Temperature dependence of the parameter $f$ for 
different quantities in 2D (left panels) and 3D (right panels) simulations.
Upper panels show results of independently fitting individual correlation
functions; lower panels show the results of constraining $\alpha$ to be equal 
for all three.}
\end{figure}

\subsection{Tail-weight parameter $f$}

Figure~\ref{showAllF} shows the temperature dependence of the parameter $f$,
the weight of the negative tail in the memory function
relative to $K_0$ (with no cut-off included).
As $T$ decreases, $f$ increases from a low value, approaching a value close to
 unity at the lowest temperatures. This increase accounts for that of the 
overall relaxation times $\tau_\alpha$ though the factor $1/(1-f)$.
The value unity cannot be reached while remaining in 
equilibrium---because this implies a diverging relaxation time---unless the
power-law behavior is cut off at the longest times. Thus there are two
possibilities for the low-temperature behavior of $f$. The first is that it
bends over and never reaches unity at finite temperature. In this case we can
avoid including a cut-off, but the 
exact temperature dependence of $f$ determines
that of $\tau_\alpha$, given that $\alpha$ seems to become 
temperature-independent at such temperatures. The second possibility is that
$f$ becomes unity at a finite temperature, for example $T\sim$0.3 in 2D or 
$T\sim$0.4
in 3D. If this happens a cut-off must be included to keep $\tau_\alpha$ finite.
When considering the full temperature range there does seem to be a bend over
towards smaller slopes at lower temperatures, but the data in  
Fig.~\ref{showAllF} are not clean enough to draw a firm conclusion about the 
limiting $T$-dependence of $f$. \footnote{We can exclude Arrhenius 
dependence  (of $1-f$), as this would imply Arrhenius dependence of 
$\tau_\alpha$}

\subsection{Comparison with KWW fits, identification of common $\alpha$}

\begin{figure}
\epsfig{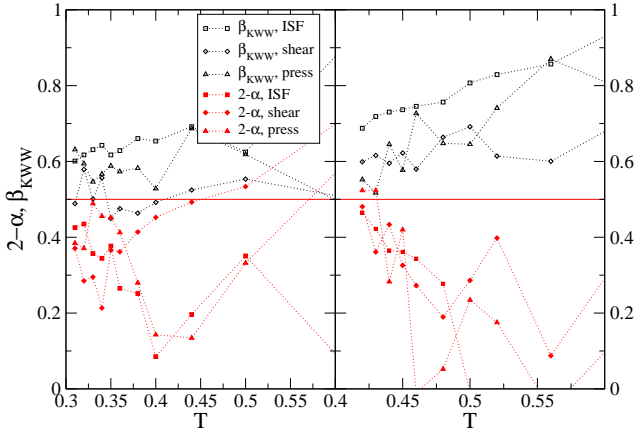}
\caption{\label{compareBetaKWW_2MinusAlpha}  (Color online)
Comparison of $2-\alpha$ 
(determined using 3-parameter, unconstrained fits) with the
 stretching parameter $\beta_{KWW}$ in a fit to the stretched
 exponential
for 2D (left) and 3D (right) simulations.. Open (black) symbols are values of
$\beta_{KWW}$, taking values above 0.5, solids (red) symbols are
values of $2-\alpha$, taking values below 0.5.} 
\end{figure}

 It is interesting to compare with a more
common characterization of the shape, the stretching parameter $\beta_{KWW}$
in the KWW function. As explained in Sec.~\ref{fEqualOneLimit}, this may be 
compared to $2-\alpha$, shown in Fig.~\ref{compareBetaKWW_2MinusAlpha}. The
values of $\beta_{KWW}$ decrease as $T$ decreases, while $2-\alpha$
increases. In a broad sense it seems that all are converging to a value around
0.5, but the degree of convergence for $\beta_{KWW}$ is much weaker.
Considering the 3D data in particular, the values of $2-\alpha$ are within
the range 0.47--0.52, 
while those of $\beta_{KWW}$ are spread over the range 0.55-0.7. The same
trend can be seen, although less clearly, in the 2D data. It is possible that
the $\beta_{KWW}$ values will converge at longer time scales, but the fact that
$\alpha$ (or $2-\alpha$) seems to converge more rapidly to a common value
suggests that this representation may be more physically meaningful. The 
value $\beta_{KWW}=$0.5, corresponding to $\alpha=$1.5, is interesting, because
it has been argued theoretically and experimentally
 that this value is generic for structural 
``alpha'' relaxation in viscous liquids, see Refs.~\onlinecite{Dyre:2005a,
Dyre:2005b, Dyre:2006b, Dyre:2007, Nielsen/others:2008}
and references therein.

\begin{figure}
\epsfig{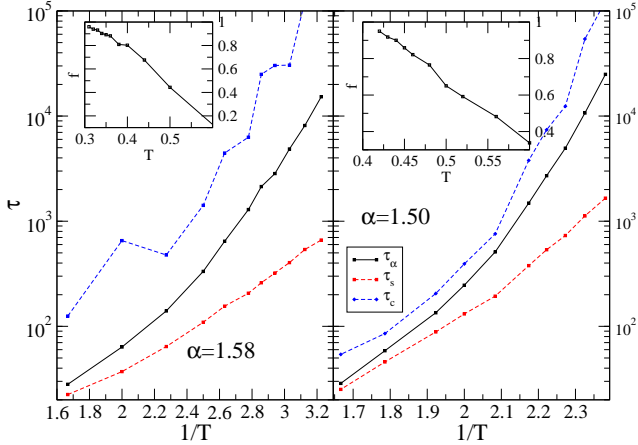}
\caption{\label{ISFtimescales_fixed_alpha}  (Color online)
Results of a constrained 4-parameter fit of $F(q_m,t)$ for all temperatures,
where $\alpha$ is held fixed. Left panel shows data for 2D, right panel for 3D.
} 
\end{figure}

\begin{figure}
\epsfig{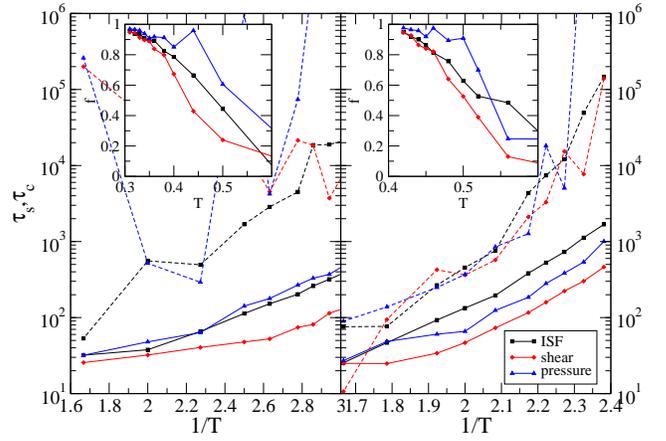}
\caption{\label{AllTimeParamsAlpha1.5} (Color online)Temperature dependence of
$\tau_s$ (symbols + solid lines) and $\tau_c$ (symbols + dashed lines)
for all correlation functions where $\alpha$ has been
fixed at 1.58 (2D, left panels) and 1.5 (3D, right panels). 
While the 2D data does not fix $\tau_c$ very
well (probably because simply it is larger relative to the relaxation time
and therefore much more subject to noise in the tail), it seems plausible that
a common value of $\tau_c$ for the three correlation functions could apply in 
3D.
} 
\end{figure}

Given the apparent convergence to a common value of $\alpha$, and the hint in
the 4-parameter fits to the ISF in 3D (lower-right
panel of Fig.~\ref{showAllAlpha}) that this parameter may in fact be 
temperature dependent, we have tried to fit the high quality ISF data with a 
common value of $\alpha$ over the whole temperature range, with the
exponential cut-off included. The results are shown in 
Fig.~\ref{ISFtimescales_fixed_alpha}. The common values of $\alpha$ that emerge
from the fit are 1.58 for the 2D system and 1.50 for the 3D system. The quality
of the ``temperature-constrained'' fits is is virtually indistinguishable by
eye from that of the unconstrained 4-parameter fits (not shown). 
The latter, are of course, numerically superior (since they are unconstrained),
 but this seems to be mostly due to better matching of the
noise in the tail. This result strengthens the case for $\alpha$ having a true
value of 1.5 in 3D, independent of temperature.
The other interesting feature of this constrained-$\alpha$
fit is that the temperature dependence of the time-scale $\tau_c$ 
of the exponential is much smoother, apparently tracking that of
the relaxation time. What about the shear and pressure data? The data was of 
too low quality for an unconstrained four-parameter fit to work, but by 
fixing $\alpha$ to be equal to 1.5 we can include the exponential cut-off, 
making it effectively a three-parameter
fit. The goodness of fit is consistently better than the original 
three-parameter fit (not shown), and now much smoother $\tau_s$ values, as well
as $\tau_c$ values are available for shear and pressure. These are shown in
Fig.~\ref{AllTimeParamsAlpha1.5}. This works better for the 3D data than for
the 2D data, which we suggest is due to the larger difference between
 $\tau_\alpha$ and $\tau_c$ in the latter case; the exponential
 cut-off has a significant effect only at longer times where the signal to
noise ratio in the correlation function is smaller, and the fit thus gives
 erratic values.

In 3D the ratio $\tau_c/\tau_\alpha$ is in the range 2--3; in 2D the
range is 8--9. Whether the cut-off time actually determines $\tau_\alpha$, or
vice versa, or whether they determine each other in a self-consistent way,
cannot be answered without a theoretical understanding and/or an explicit model
for the relaxation. Nor do we know what may be the cause of the different 
ratios in 2D and 3D. To
get a clearer picture more systems should be analyzed, and the possibility that
$\tau_c$ is size dependent cannot be excluded.

\subsection{Comparison of the fitted $K_0$ correspond with the actual short-time 
decay of the inherent quantities}

A feature of our fitting procedure is that we leave the zero-time value of
the correlation function as a fitting parameter, since the actual zero-time
value of the true correlation function includes contributions from vibrational
motion which rapidly decays. It is worth asking whether the value returned by 
the fit coincides with the value obtained by a careful determination of the
inherent correlation function. Since we focus on time 
intervals of order $\Delta t\sim20$, we can run relatively long simulations,
and obtain good statistics for inherent quantities, by minimizing at similar 
intervals. For example minimizing once every 1000 time 
steps (100 times less frequently than before)
corresponds to an interval $\Delta t=10$, and means that the simulation
time is not dominated as much by the cost of minimizing.
This was done in both the 2D and 3D systems for two temperatures:
0.34 and 0.40 for 2D and 0.44 and 0.52 for 3D. In addition runs with 
minimization
every 100 time steps were carried out for 2D, $T=0.34$.

\begin{figure}
\epsfig{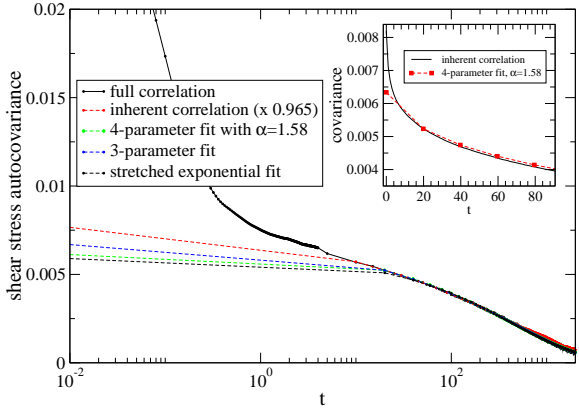}
\caption{\label{compareShortTimeDecorrelation2DT0.34} (Color online)
Shear stress correlation 
functions and fits for 2D system at $T=0.34$. Main figure, comparison of true, 
inherent and fitted functions. The inherent function was based on minimization 
at intervals of 10 (1000 time steps); it has been rescaled to match the full
 correlation at $t=20$ (by a factor 0.97). The inset shows a zoom in of the 
inherent correlation function and the 4-parameter fit with $\alpha=1.58$ at 
short times. In this case the inherent correlation function was based on
 minimization every 1 unit (100 time
steps); here the rescale factor is 0.99. The need to rescale simply reflects
the sampling error in the variance ($t=0$ value); that of the autocovariance
is almost the same (for relatively short lags), hence different sampling 
results in curves which are more or less proportional to each other. When 
comparing normalized correlation at, say $t=20$, the difference between
minimization every 10 time steps (relatively short simulation times)
and every 1000 time steps (relatively long simulation times)
is of order 1\%.}
\end{figure}

\begin{table}
\begin{tabular}{|c|c|c|c|c|c|c|}
\hline
system & $T$  & corr. & $\psi(\Delta t)_{I}$ & 
$\psi(\Delta t)_{1.5}$ & $\psi(\Delta t)_{3-p}$ & $\psi(\Delta t)_{SE}$ \\
\hline
2D     & 0.34 & ISF   & 0.86 & 0.94 & 0.93 & 0.95 \\
2D     & 0.34 & shear & 0.64 & 0.83 & 0.74 & 0.84 \\
2D     & 0.34 & press.& 0.90 & 0.95 & 0.95 & 0.95 \\
2D     & 0.40 & ISF   & 0.72 & 0.83 & 0.71 & 0.84 \\
2D     & 0.40 & shear & 0.46 & 0.56 & 0.60 & 0.54 \\
2D     & 0.40 & press & 0.77 & 0.85 & 0.70 & 0.82 \\
3D     & 0.44 & ISF   & 0.94 & 0.97 & 0.97 & 0.98 \\
3D     & 0.44 & shear & 0.78 & 0.91 & 0.90 & 0.89 \\
3D     & 0.44 & press.& 0.91 & 0.94 & 0.93 & 0.96 \\
3D     & 0.52 & ISF   & 0.70 & 0.79 & 0.73 & 0.81 \\
3D     & 0.52 & shear & 0.34 & 0.43 & 0.46 & 0.40 \\
3D     & 0.52 & press.& 0.65 & 0.68 & 0.66 & 0.73 \\
\hline
\end{tabular}
\caption{\label{shortTimeDecorrelation}Value of correlation functions 
(normalized to equal unity at time zero) at time 
$\Delta t=20$; comparison between actual inherent value (``I'') determined from
the autocorrelation function of inherent quantities and values inferred from 
fitting the correlation function at longer times. Values for three fits are 
shown: (``1.5'') the fit with $\alpha$ fixed at 
be 1.5, including the exponential cut-off; (``$3-p$'') the 3-parameter fit with
 no cut-off; and (``SE'') 
the stretched exponential fit. The correlation at $\Delta t$ implied by the
fits is higher than the actual correlation; alternatively put, the amount of
decorrelation that takes place on the time scale $\Delta t$ is less than would
be expected from the fits.}
\end{table}

Table~\ref{shortTimeDecorrelation} shows values of the normalized correlation
function after one time interval $\Delta t=20$, for two selected temperatures
in each system, determined directly from the inherent state dynamics and
inferred from different fits to the data for $t\geq\Delta t$. Although there 
are some exceptions (particularly for 2D, $T=0.40$), the trend is that the 
value of the normalized inherent
 correlation implied by the fits is higher than the actual value. 
For example for 2D at $T=0.34$, considering the intermediate scattering 
function, the actual correlation is 0.86, while the fits all imply a value 
between 0.93--0.94. Another way of putting this is that the amount of {\em 
decorrelation} that takes place between time zero and time $\Delta t$ is 
greater than is implied by the fits of the correlation at longer times (note
that $\psi(\Delta t)=\psi_1=1-K_0$). This 
can be thought of as somewhat analogous to the extra relaxation associated 
with the vibrational motion which has been removed, but this is part of
the inherent relaxation. Schr{\o}der et al. also noticed a short-time 
component of the inherent structural 
relaxation (intermediate scattering function) which was not accounted for by 
stretched exponential fits.\cite{Schroder/others:2000} This was 
particularly noticeable at high temperatures and appeared to disappear in the
activated regime. The comparison above suggests that this extra relaxation does
not vanish, even at quite low temperatures. This discrepancy, between the very
short-time relaxation and that for later times ($t>\Delta t$),
presumably corresponds to the change of apparent power-law exponent in the
memory function around this time scale (section~\ref{MatchingCorr}).
One could speculate that the existence of this extra component of the
inherent relaxation could be associated with very small energy barriers, much
smaller than the temperature. Such barriers would separate distinct inherent
states, and thus ``transitions'' would show up in the inherent correlation 
functions, but these transitions would not be activated. The author has
determined that the energy barriers in these systems are exponentially 
distributed, thus there are always some small barriers.\cite{Bailey:2009a}

\subsection{Wavenumber dependence of intermediate scattering 
function}


\begin{figure}
\epsfig{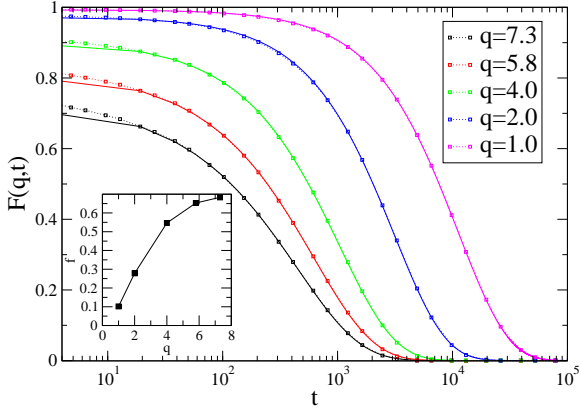}
\caption{\label{qDepFits3DT0.48} (Color online)
 Wavenumber-dependent intermediate scattering
function (symbols and dotted lines) and 4-parameter fits with $\alpha=1.5$ 
(solid lines), for 3D system, $T$=0.48. The inset shows the $q$-dependence of 
the tail-weight parameter $f$.}
\end{figure}

For simplicity, in studying the intermediate scattering above, we have 
restricted the analysis to a particular wavenumber, $q_m$, and to 
large particles. Given that a common value of the exponent $\alpha$ 
seems to be valid for the three correlation functions we have focused on so 
far, it makes sense to ask whether this applies for the intermediate scattering
function at other $q$-values. In particular, what happens at low $q$, where we
know that the relaxation becomes more exponential? Fig.~\ref{qDepFits3DT0.48}
shows data and fits for 3D at $T=0.46$. Here 4-parameter fits with fixed
 $\alpha=1.5$ were used.

\begin{figure}
\epsfig{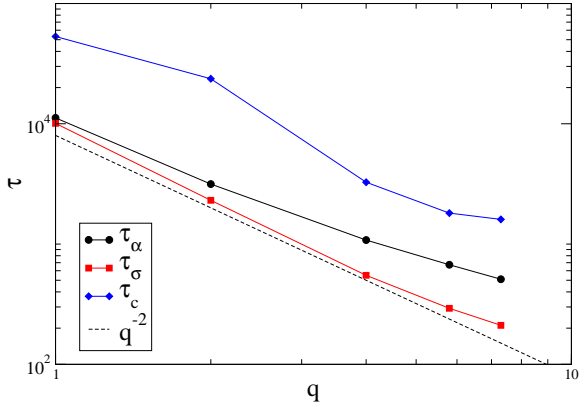}
\caption{\label{qDep3DT0.48TimesAndF} (Color online)
 Main figure, double-log plot of the 
$q-$dependence of the time-scales associated with the 4-parameter fit: the
relaxation time $\tau_\alpha$, the inverse of short-time rate, $\tau_s$,
and the exponential cut-off time $\tau_c$. A line proportional to $q^{-2}$ is
shown as a guide to the eye.
 The $q$ values used are as in Fig.~\ref{qDepFits3DT0.48}.}
\end{figure}

Figure~\ref{qDep3DT0.48TimesAndF} shows the $q$-dependence of the relaxation
time $\tau_\alpha$, inverse of the short time rate, $\tau_s$ and the cut-off
time of the memory function $\tau_s$. The inset shows the $q$-dependence of 
$f$. While the power-law exponent is fixed at 1.5 for all $q$-values, $f$
varies strongly, approaching zero at small $q$. This corresponds to
relaxation that becomes exponential in this limit. Consistent with this,
$\tau_s$ and $\tau_\alpha$ approach each other. $\tau_c$ seems to follow 
$\tau_\alpha$. In normal diffusion one expects the relaxation time to be
 proportional to $q^{-2}$, indicated by the dashed line. The relaxation time
seems to approach this dependence in the low-$q$ limit (as is expected, given
that diffusion should be normal at long length and time scales). More 
interesting is that $\tau_s$ seems to show an inverse square dependence over 
most of its range, corresponding to normal-diffusive behavior
being observed at {\em short} times. This will be discussed below.

\section{\label{Discussion}Discussion}

\subsection{Generalized Langevin equations and 
continuous-time random walks}

What does a negative inverse power-law memory function mean? One interpretation
derives from the formalism of the generalized Langevin equation
(GLE), which offers a framework for interpreting memory functions. The
GLE for a dynamical variable $A(t)$ is a stochastic differential equation

\begin{equation}
\frac{d A}{dt}  = -\int_0^{t} K(t-t')A(t')dt' + F(t).
\end{equation}

The first term on the right hand side
involves an effective friction, non-local in
time, where the memory kernel $K(t-t')$ connects the friction at current time
$t$ with the value of $A$ at a (previous) time $t'$. The last term 
$F(t)$ is the so-called random force, or noise term. According to the second 
fluctuation-dissipation theorem, 
the memory kernel is the autocorrelation function of the noise, or

\begin{equation}
K(t) = \angleb{F(0)F(t)}.
\end{equation}

\nod It may also be shown that this memory kernel $K(t)$ is the memory function
associated with the autocorrelation of $A(t)$---the same memory function used
in the analysis of this paper. 
Therefore one can interpret the obtained memory functions 
as the autocorrelation functions of the random force of the corresponding
GLE. By ``force'' is meant the stochastic increment to the quantity $A$; in the
original version of the Langevin equation $A$ was the velocity of a particle,
whose change was indeed a force (divided by mass).
Negative $K(t)$ for $t>0$ implies that the changes, which are 
presumably related to the so-called ``flow-events'' which define the long-time
dynamics, are anti-correlated. The simplest interpretation of this would be
forward-backward correlation of events---that is, 
an event is more likely to be the reverse of the
previous one than to be something else. The parameter $f$ is then a
direct measure of this anti-correlation. As it approaches unity, this 
tendency becomes so strong that the relaxation time diverges (see the next
subsection).

An alternative interpretation involves
uncorrelated events but with a non-exponential waiting-time distribution. This
can be understood within the continuous-time random walk formalism of
Montroll and Weiss.\cite{Montroll/Weiss:1965} The Montroll-Weiss equation
relates the Laplace transform of $F(q,t)$ to
the Laplace and Fourier transforms of the waiting-time and jump-size
distributions, $\tilde p_w(s)$ and $\tilde p_j(k)$, respectively. It is 
straightforward to use it to derive
a relation between the waiting-time distribution and the memory function.
Skipping the details, this yields

\begin{equation}
\tilde p_w(s) = \frac{\tilde K(s)}{\tilde K(s) + s(1-\tilde p_j(k))}.
\end{equation}

This relation can be considered a formal equivalence of the two 
pictures---anti-correlated random forces and non-exponential waiting-time
distribution. The question of which picture is {\em physically} more 
relevant can only be determined by more detailed analysis of the dynamics
in simulations. The existence of forward-backward correlations has indeed 
been noted by Heuer and coworkers 
(see Ref.~\onlinecite{Heuer:2008} for a review of this work), but 
they argue that these are trivial, operating only on relatively
short time scales.
They have proposed a procedure for coarse-graining the dynamics by grouping
the basins of attraction of inherent states together to form ``meta-basins'' 
(MBs). The definition of an MB is based on the requirement that once the 
system has escaped from an MB, it is unlikely to return there.
They have shown that the dynamics on long time scales is
essentially a random walk among MBs, but with non-trivial
waiting-time distribution.\cite{Rubner/Heuer:2008} If we accept this 
interpretation, the memory function analysis presented here could be used as
a way to find the waiting-time distribution from the correlation function. For
example, in the $f\rightarrow1$ limit with no cut-off (see next section) one
finds that the waiting time distribution has a power-law tail with the
exponent $\alpha$ as the memory function itself.

On the other hand, is it possible that the equivalence between the waiting time
distribution point of view and the forward-backward point of view is in fact 
more than formal? The non-exponential distribution of waiting times
could seen as a consequence of forward-backward correlations.
The size of an MB is not a fixed  property of the 
landscape,\cite{Doliwa/Heuer:2003b} instead being defined in terms of dynamical
behavior: a given MB
can increase in size as temperature decreases, reflecting the increased 
tendency for returns (the system needs to get further away on average 
before its chance of escaping for good exceeds 50\%). Correspondingly, the
``anti-correlation tendency'' $f$ increases towards unity; as long as it 
is less than unity, however, a diffusive regime is reached---the system will
eventually forget where it started. In the ``CTRW picture'', as 
long as the waiting-time distribution
has a finite mean, upon sufficient coarse-graining it will become exponential
and therefore the dynamics will be diffusive in the normal sense.
The point of Heuer and co-workers is that there is a coarse-graining time scale
at which events are uncorrelated but the waiting-time distribution is 
non-trivial (non-exponential). It is also true in the ``GLE picture'' 
involving anti-correlated events that, as long as $f<1$ or there is an 
exponential cut-off, a normal diffusive regime must also eventually be reached 
under coarse-graining in time. But it is not clear yet whether 
coarse-graining in the GLE picture can produce a dynamics that corresponds to
the CTRW picture. This is a matter for future investigations.

\subsection{\label{fEqualOneLimit}The $f=1$ limit: sub-diffusion}

In this section we discuss briefly the limit $f\rightarrow 1$, in the case
where there is no exponential cut-off. As mentioned,
in this limit the relaxation time diverges, but the dynamics defined by this
memory function are still meaningful, and have been studied in the context of
anomalous diffusion, fractional Brownian motion and related processes. 
\cite{Metzler/Klafter:2000, Rangarajan/Ding:2003} Applications include
protein dynamics\cite{Kneller/Hinsen:2004, Min/others:2005} and dielectric 
response of glassy media. \cite{Goychuk:2007} 
A noise term in a GLE with this kind of 
correlation function is termed fractional Gaussian noise; when it acts on 
an unbound degree of freedom, the latter undergoes fractional Brownian
motion. Correlation functions generally involve a function known as the
Mittag-Leffler (ML) function:\footnote{The argument of the Mittag-Leffler
function is actually $(t/\tau_{ML})^\beta$ for a time scale $\tau_{ML}$; so 
the correlation function should really be called a ``stretched Mittag-Leffler
function'', but it seems to be standard to refer to the correlation function
also as a Mittag-Leffler function, which is not strictly correct.}
They behave at intermediate times like a stretched exponential, with stretching
parameter $\beta_{KWW} = 2-\alpha$, before
switching to a power-law at long times, such that the integral is infinite.
This is why it made sense  to compare $2-\alpha$ with the fitted $\beta_{KWW}$
in Fig.~\ref{compareBetaKWW_2MinusAlpha}.

There are two broad classes corresponding to positively or negatively
correlated noise.\footnote{These can also be defined in terms of the so-called
Hurst exponent $H$ being respectively greater than $\half$, which also 
corresponds to $0<\alpha<1$ or less than $\half$, corresponding to 
$1<\alpha<2$. At $H=\half$ the exponent is formally equal to unity but its
coefficient vanishes thus white (uncorrelated) noise is recovered.}
When comparing to the literature on
fractional diffusion and related processes it is important to realize that
the Langevin equation is typically formulated with the dynamical variable
 $A(t)$ as the velocity of a particle. For a free particle, if the noise is 
fractional Gaussian with negative correlation, 
then the velocity autocorrelation function
is an ML function. Its integral diverges, therefore the position dynamics 
is actually super-diffusive. Applications more typically involve
sub-diffusive motion of a harmonically 
bound particle.\cite{Min/others:2005} The noise acting on the velocity
is positively correlated and it is the position autocorrelation function that 
decays as an ML function.

The memory function corresponding to this limit is $\tilde K(s)=s^{\alpha-1}$,
which is clearly scale invariant: Rescaling, or coarse-graining in time 
produces the same dynamics. This corresponds to a fixed point in the
renormalization group sense. For negatively correlated noise, the fixed point
is unstable (see the chapter by Qian in 
Ref.~\onlinecite{Rangarajan/Ding:2003})---if we are not quite in this limit, 
then rescaling will eventually takes us away from it and back to the normal
diffusive regime, as discussed in the previous subsection. The results of the
analysis presented in this paper could be therefore interpreted as saying
that glassy dynamics approaches sub-diffusive dynamics with a
particular exponent $\alpha$, which (in 3D) is very close to 1.5. As the limit
$f=1$ is approached the relaxation becomes close to a
 Mittag-Leffler function, resembling a stretched
exponential with $\beta_{KWW}=0.5$. As we have seen, even at the lowest
temperatures, actually fitting with a stretched exponential yields a different 
exponent, so it is hard to see the convergence to 0.5, while this is clearer
in the memory function based approach. The theoretical challenge is, of course,
to understand where this exponent comes from, although there are models
for viscous liquid dynamics which predict it (Refs.~\onlinecite{Dyre:2005a,
Dyre:2005b, Dyre:2006b, Dyre:2007, Nielsen/others:2008} and references within).

Does this limit ever get reached? Although $f$ clearly approaches unity, it has
not reached it yet at the temperatures simulated so far. Moreover, there is
always an exponential cut-off in the memory function. Thus regardless of 
whether $f$ actually reaches unity or not, it will at some point be the
exponential cut-off which will determine the relaxation time. Looking at
Fig.~\ref{AllTimeParamsAlpha1.5}, for example, suggests that this will happen
around $T=0.40$ in 3D. Simple-minded linear or quadratic extrapolation of $f$ 
suggests this also. If we set $f=1$ for temperatures below say $T=0.40$ and
assume that the exponential cut-off, $\tau_c$ is a fixed multiple of 
$\tau_\alpha$ (actually the ratio seems to increase), then one can make an
extrapolation of $\tau_\alpha$, which gives a sharp crossover to an almost
Arrhenius form (not shown). If $f$ approaches unity more smoothly, then the
crossover will not be so abrupt, but in any case
 we predict that a change in the 
temperature dependence will occur around $T=0.40$.

\subsection{Short-time dynamics}

The parameter $\tau_s=\Delta t/K_0$  is a measure of the dynamics at short
times. By fixing the time scale, it measures how much activity happens in a
given time, as opposed to standard measures of relaxation, which concern
how much time is required for a given amount of relaxation to occur. For
exponential relaxation there is no difference, for relatively small intervals
(or long relaxation time $\tau_\alpha$): 
If the interval is $\Delta t$, then the amount of correlation at this time is 

\begin{equation}
1-K_0 = \exp(-\Delta t/\tau_\alpha)\simeq 1- \Delta t/\tau_\alpha,
\end{equation}

\nod which gives $\tau_s=\tau_\alpha$. Thus a difference between these two
quantities is equivalent to non-exponential relaxation.  When $\Delta t$ is 
much smaller than $\tau_\alpha$, the expected number of activated events that 
take place in a small volume in a time interval of this length is much smaller 
than unity, thus we cannot expect to see any correlation effects.
We suggest therefore that
$\tau_s$ is like a ``bare'' relaxation time, measuring the dynamics before
correlation effects play a role. Recall
 (Fig.~\ref{compareShortTimeDecorrelation2DT0.34}) that there is a noticeable
amount of (inherent) relaxation in this time interval which is not accounted 
for by the fitted $\tau_s$. We assume this to be non-activated relaxation,
and not relevant for what takes place at longer time scales (in this sense it
resembles the vibrational part of the relaxation). The evidence supporting the 
idea of $\tau_s$ as a measure of the bare, uncorrelated relaxation rate is the 
near-Arrhenius temperature dependence (see Fig.~\ref{AllTimeParamsAlpha1.5}), 
and the near $q^{-2}$ wavenumber dependence (Fig.~\ref{qDep3DT0.48TimesAndF})
The use of a fixed, relatively
short time interval to extract an apparent ``underlying Arrhenius behavior'' 
in viscous liquid dynamics was recently demonstrated by de Souza and Wales
\cite{deSouza/Wales:2006a, deSouza/Wales:2006b}. They studied the diffusion
constant determined by the mean squared displacement in a fixed time interval
of 25 LJ units. They were also able to explicitly relate the deviations from
exponential relaxation  to anti-correlation of events in subsequent time 
windows. The present results are essentially equivalent, but obtained in
the more general context of studying an arbitrary autocorrelation function.

\section{Conclusion}

We have presented a method for fitting the kind of slowly 
decaying autocorrelation functions typical of the dynamics of 
glass-forming liquids. An explicit functional form is postulated, not for the
autocorrelation function, but for its associated memory function: a positive
zero-time parameter $K_0$, and a negative inverse 
power law for non-zero time. An long-time exponential cut-off may optionally
be included. This involves another parameter, although we have shown that good
fits may be obtained by fixing the exponent to a common value, reducing the 
number of parameters per data set essentially to three. While it is
possible to obtain the exact memory function from the autocorrelation function,
it is not useful to fit the former directly. Rather we match the 
autocorrelation function by optimizing the parameters defining the 
memory function. Using a quite general functional form for $K(t)$ we noted
a crossover from one apparent power-law to another at relatively short times,
which allowed a simplification by considering data separated at times longer
than the cross-over, namely the simple power-law form (with possible 
exponential cut-off). In particular we chose $\Delta t=20$ in Lennard-Jones
units for all of the analysis. While the fits are about as good as fits with a
stretched exponential, the parameters obtained 
have arguably greater physical significance than those of the latter. In
particular the exponent $\alpha$ seems to be more or less independent of the
relaxing quantity, and indeed of temperature (as long as the exponential 
cut-off is included), which is not the case for the stretching parameter 
$\beta_{KWW}$. Moreover the parameter $K_0$ may be interpreted as a short time
rate; its temperature dependence is, interesting, significantly weaker than
that of the alpha relaxation times for the different autocorrelation functions.
As temperature decreases, the amplitude of the power tends to approach a
limiting value associated with the mathematical description of anomalous 
diffusion/relaxation characterized mathematically by the Mittag-Leffler
function. We hypothesize that this value is reached at finite temperature
but the associated divergence of relaxation time is removed by the
presence of an exponential cut-off. Including the latter in the fits for the
self-intermediate scattering function allowed the identification of a third
time scale, longer than the alpha time.
It is hypothesized that the strong non-Arrhenius temperature dependence is
due to a crossover from the time associated with the short-time rate to that
associated with the long-time cut-off of the memory function, and that in 
particular at temperatures somewhat lower than those simulated, the 
non-Arrhenius behavior will weaken noticeably.

\begin{acknowledgments}
The centre for viscous liquid dynamics ``Glass and Time'' is sponsored by the
 Danish National Research Foundation (DNRF).
\end{acknowledgments}

\appendix

\section{Memory function fitting}

Here we describe in a little more detail how we optimize the fit. 
We start by describing the procedure using the general fitting form
which includes the piecewise linear function $L(t)$.
Suppose first we have a correlation function based on inherent dynamics, so
that there is no vibrational decay at short times, and in particular the
$t=0$ value is known. Then we may normalize by the latter to get the
normalized autocorrelation function $\psi(t)=\psi_n$ where $n=t/(\Delta t)$
$n$ is the integer corresponding to $t$. We write the fitting form as 
$K^{fit}_n(\{p_a\})$, where $p_1, p_2, \ldots$ are the parameters of the fitting
form. The node values $L_k$ are not used directly as parameters; rather their
logarithms are. This constrains them to be positive and minimizes 
numerical problems associated with the broad range of values that they turn
out to have. For given values of the $\{p_a\}$, we can easily compute the 
associated correlation function $\psi^{fit}_n(\{p_a\})$ and then
the objective function

\begin{equation}
F_{obj}(\{p_a\}) = \sum_n \left(\psi(t) -  \psi_{gen}(t,\{p_a\})\right)^2.
\end{equation}

\nod To minimize $F_{obj}$ we used the conjugate gradient
technique,\cite{Press/others:1987} calculating the gradient numerically.

For the main data analysis we switched to using simpler functional forms
without the piecewise linear function. Also significant is that we used only
the true dynamical correlation functions (as opposed to those of the
inherent dynamics) which are more accurate and smooth; 
this removed the need to do expensive minimizations. By 
considering times at intervals of order 20, we could avoid the vibrational
decay, except that now the correct initial value of correlation function had to
be treated as an unknown parameter. Since this affects the normalization,
it is important to handle this carefully. Suppose we have a non-normalized
correlation function at discrete times, not including time
zero: $C_1, C_2, C_3, \ldots$. Since we are going to normalize anyway, we choose
as the parameter not $C_0$, but the ratio $R_{01}\equiv C_0/C_1$. And since this
is required be greater than unity, it is represented internally (i.e., within 
the fitting algorithm) in terms of a logarithm $\theta_R$: $R_{01} = 1 + 
\exp(\theta_R)$, where $\theta_R$ is the actual parameter
varied by the algorithm. Note that $R_{01}$ is directly related to $K_0$:

\begin{equation}
K_0 = 1-\psi_1 = 1-(C_1/C_0) = 1-1/R,
\end{equation}

\nod so that one may equivalently consider $K_0$ as the fitting parameter.
An important technicality must be mentioned here. If the objective function
is defined as above, 
there is a problem due to the fact that the normalization factor
is now an adjustable parameter. This means that the objective function could be
reduced by making $R_{01}$, and hence $C_0$, larger. This was found to be a 
problem for data sets from higher temperatures where the relaxation times were
relatively short and therefore the data was not sufficient to constrain $R_{01}$
to reasonable values. The problem was solved by normalizing by the fixed 
quantity $C_1$ instead (but only for the purposes of defining the objective
 function). For $n>0$ (times greater than $\Delta t=20$) 
we used in some cases the single power law of Eq.~\ref{power_law_Km}.
The tail-weight fraction $f$ is constrained to lie between 0 and 1 by 
expressing it internally as the $f=\arctan(f_1)$ where $f_1$
 is unconstrained. The exponent $\alpha$ was represented as itself 
because it tended to vary within a relatively small range (1.5--2.2),
 with no risk
of assuming values which would cause numerical problems. In other cases an
exponential cut-off factor was included, Eq.~\ref{exp_power_lawKm},
 which involves the timescale $\tau_{cut}$. Because
this can vary over several orders of magnitude, to aid convergence
it was expressed internally in terms of its 
logarithm, $\tau_{cut}=\exp(10\theta_\tau)$, where the factor 10 was found to 
improve convergence (presumably by making the typical values and range of 
$\theta_\tau$ comparable to those of other parameters).

\end{document}